\documentclass[seceq]{ptptex}

\usepackage{graphicx}
\title{%        %You can use \\ for explicit line-break.
Quarkonium production in PbPb collisions at the LHC%
}

\author{%       %Use \scshape for the family name.
Rapha\"el \textsc{Granier de Cassagnac}, \\ on behalf of the ALICE, ATLAS and CMS collaborations%
}

\inst{%     %Affiliation, neglected when [addenda] or [errata].
Laboratoire Leprince-Ringuet, \'Ecole Polytechnique, IN2P3-CNRS, France
}

\abst{%         %This abstract is neglected when [addenda] or [errata].
I review the recent quarkonium measurements performed at the LHC in PbPb collisions at $\sqrt{s_{NN}}=2.76$~TeV.
}

\begin{document}

\maketitle

\section{Introduction}

Quarkonium suppression, as it was introduced by Matsui and Satz in 1986~\cite{MatsuiSatz}, should be a direct signature of deconfinement. Indeed, these heavy quark-antiquark bound states, the lighter of which being the $J/\psi$ particle, should melt in the quark gluon plasma thanks to the weakening of the colour force. Moreover, various bound states with different binding energies should melt at various temperatures. Measuring different quarkonia ($J/\psi$, $\psi'$, $\chi_c$, the $\Upsilon$ and $\chi_b$ families) could thus serve as a thermometer of the produced medium, each quarkonium suddenly disappearing above its proper dissociation temperature~\cite{Mocsy}. The most tightly bound state is believed to be the $\Upsilon(1S)$ and it should thus be the last to disappear.

However, both at SPS and RHIC, it was soon realized that while the hot quark gluon plasma could indeed melt quarkonia, cold normal nuclear matter can also absorb them. At RHIC energies, namely $\sqrt{s_{NN}} = 200$~GeV, measurements of $J/\psi$ suppression in AuAu collisions by the PHENIX experiment~\cite{Phenix1,Phenix2} brought up two surprises. First, at midrapidity, the amount of suppression is surprisingly similar to the one observed at SPS if plotted as a function of the number of participants $N_{part}$. There is no fundamental reason for this to happen, since the energy density should be higher at RHIC, and the cold nuclear effects could be drastically different. Second and even more surprising is the fact that, at forward rapidity, $J/\psi$ are further suppressed (by approximately 40\%).

These RHIC \emph{puzzles} can be qualitatively explained by two effects which are difficult to distinguish: 1) gluon shadowing or saturation could further reduce the yield at forward rapidity, 2) $J/\psi$ could be recreated, mostly at midrapidity, by recombination of uncorrelated $c$ and $\overline{c}$ quarks.

Another interesting observation at RHIC is the fact that higher transverse momentum ($p_T$) $J/\psi$ are less suppressed, as reported by the STAR experiment \cite{Star1,Star2}.

In December 2010, the Large Hadron Collider delivered the first heavy-ion collisions in the TeV regime, namely about 9~$\mu$b$^{-1}$ of PbPb collisions at $\sqrt{s_{NN}}=2.76$~TeV. In March 2011, about 200 nb$^{-1}$ of pp collisions at the same energy were also delivered. First quarkonium results were soon presented by the ALICE, ATLAS and CMS collaborations. A description of the three experiments can be found in Ref. \cite{AliceAp, ATLASAp, CMSAp}

\section{High $p_T$ J/$\psi$}

\begin{figure}[hbtp]
  \begin{center}
    \includegraphics[width=0.45\linewidth]{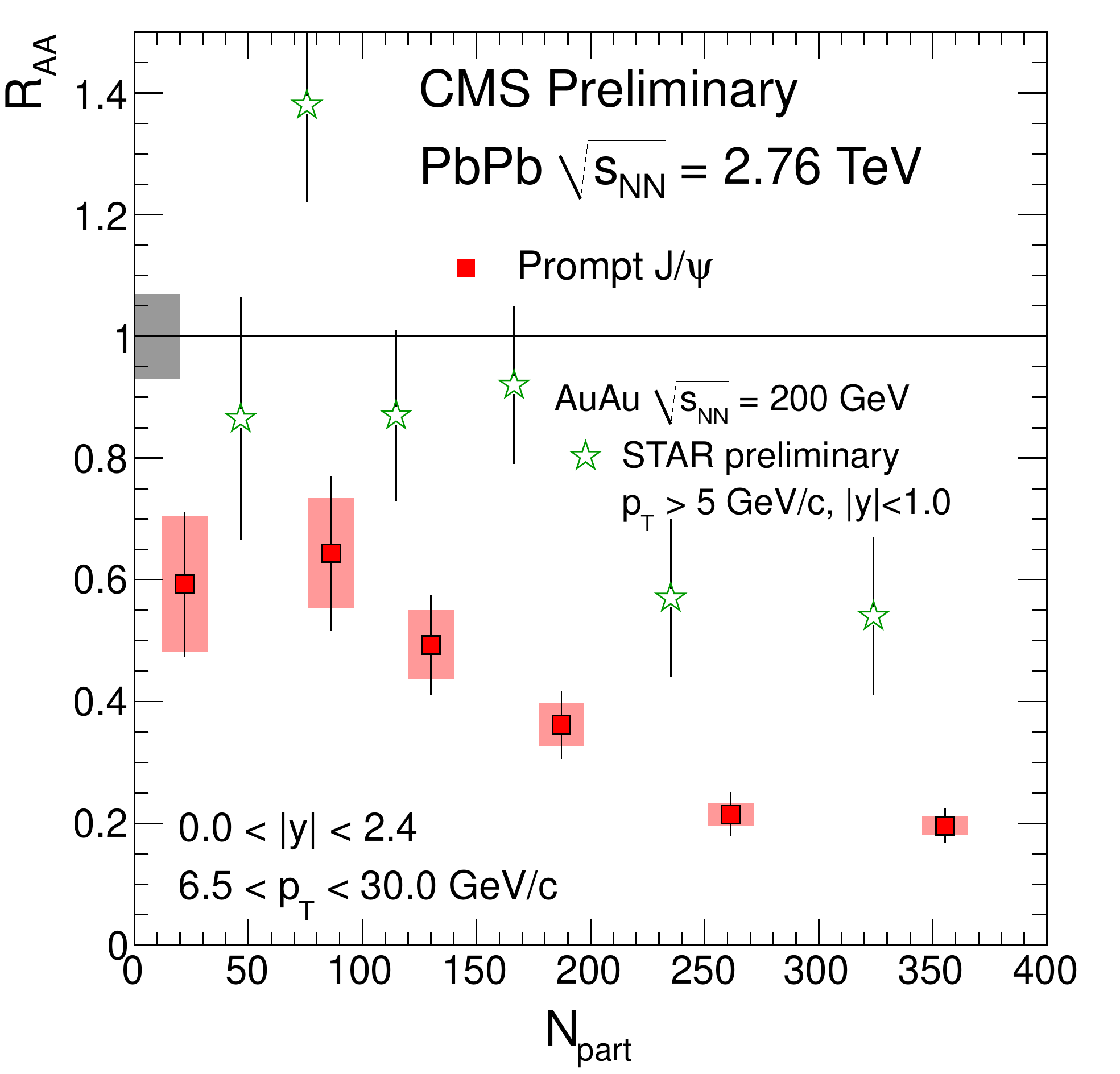}
    \includegraphics[width=0.45\linewidth]{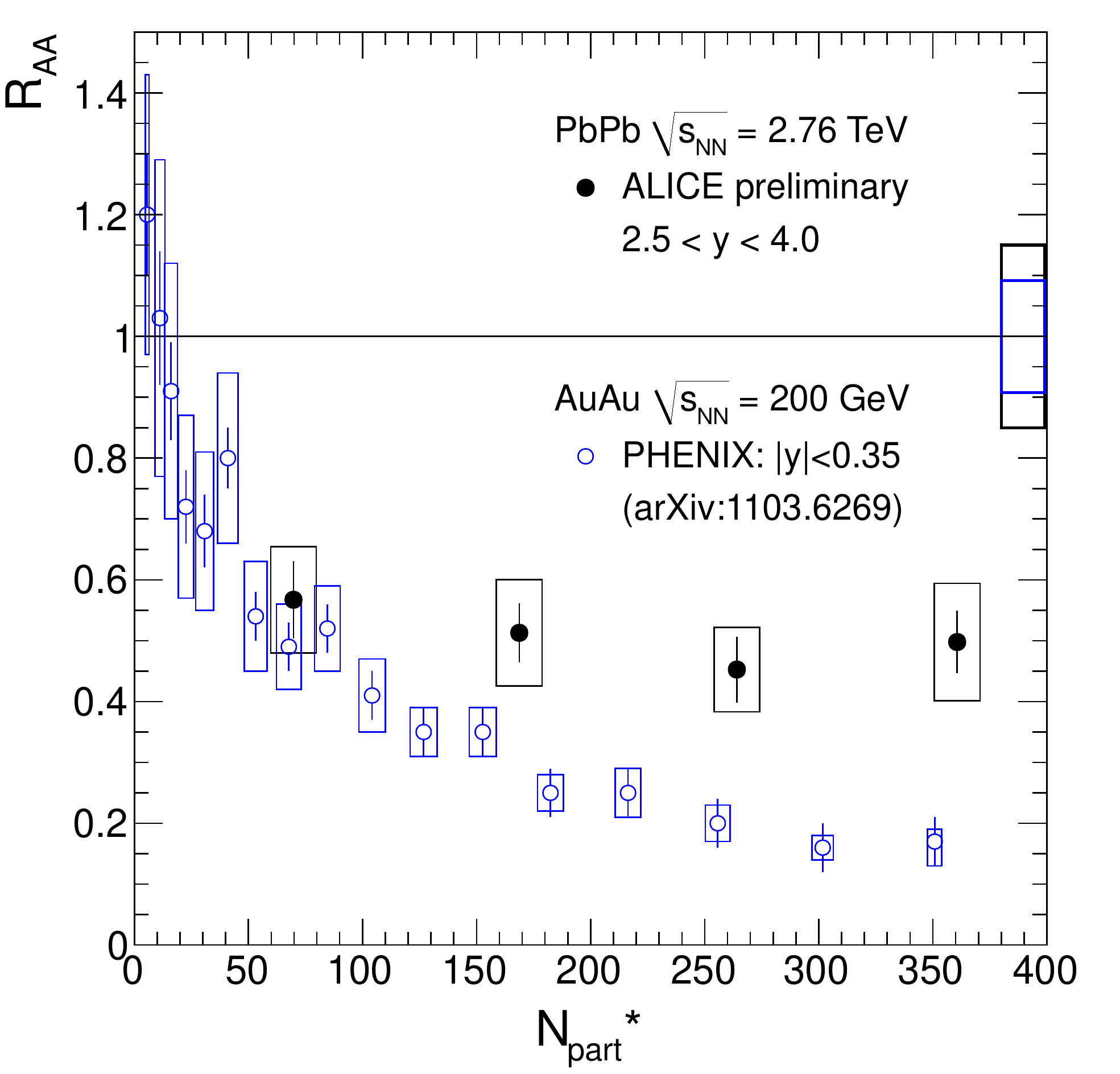}
    \caption{Nuclear modification factor as a function of the number of participants. Left: high-$p_T$ prompt $J/\psi$ from CMS compared to STAR. Right: low-$p_T$ $J/\psi$ from ALICE compared to PHENIX.}
    \label{fig:jpsi}
  \end{center}
\end{figure}

ATLAS first published a strong suppression of the $J/\psi$ yield by comparing the central and peripheral (40$-$80\%) collisions\cite{Atlas}. The acceptance of the high-field multi-purpose experiments is best suited for $J/\psi$ towards larger $p_T$ and 85\% of the $J/\psi$ reported by ATLAS have $p_T > 6.5$~GeV/$c$. The CMS collaboration later reported a preliminary analysis~\cite{CMSQuarkonia} of the nuclear modification factor ($R_{AA}$), by comparing to the pp data taken at the same energy. In this analysis, prompt $J/\psi$ (directly produced or coming from decays of $\psi'$ or $\chi_c$) and non-prompt $J/\psi$ (coming from $B$-meson decays) are separated for the first time in heavy-ion collisions. The results for prompt $J/\psi$ are reported in the left panel of Fig.~\ref{fig:jpsi}. The analysis is restricted to $p_T > 6.5$~GeV/$c$ and compared to results from the STAR collaboration\cite{Star2} for $p_T > 5$~GeV/$c$. In this high-$p_T$ kinematical range, $J/\psi$ appear to be more suppressed at LHC than at RHIC. The $p_T$ and rapidity dependencies are also investigated and a slightly lower suppression is seen at forward rapidity, see details in Ref.~\cite{CMSQuarkonia}

\section{Low $p_T$ J/$\psi$}

With its lower magnetic field and forward muon absorber and spectrometer (covering $2.5<y<4.0$), the ALICE apparatus is well suited to measure $J/\psi$ down to the lowest $p_T$. Preliminary results\cite{Alice} of the inclusive (in $p_T$) nuclear modification factor were released and are reproduced on the right panel of Fig.~\ref{fig:jpsi}, as a function of $N_{part}$\footnote{For ALICE, the number of participants is reweighted according to the number of elementary collisions ($N_{coll}$).}, together with the PHENIX measurement at forward rapidity ($1.2<|y|<2.2$). The contribution from $B$-meson decays is not subtracted in both cases but should be of the order of 10\% for ALICE, as measured on pp collisions, and less at RHIC energy. In this low-$p_T$ range (but in slightly different rapidity range), $J/\psi$ appear to be less suppressed at LHC than at RHIC.

A possible explanation of this opposite trend would be the turn-on of regeneration processes, which should be more important for lower $p_T$. Indeed, multi-component models including regeneration~\cite{ZhaoRapp} tend to be able to reproduce the observed features of the low and high-$p_T$ $J/\psi$ LHC data. ALICE also reported a promising analysis~\cite{Alice} in the dielectron channel at mid-rapidity ($|y|<0.8$) but the available statistics only allow two bins in centrality and large uncertainties for now.

\section{The $\Upsilon$ family}

\begin{figure}[hbtp]
  \begin{center}
    \includegraphics[width=0.38\linewidth]{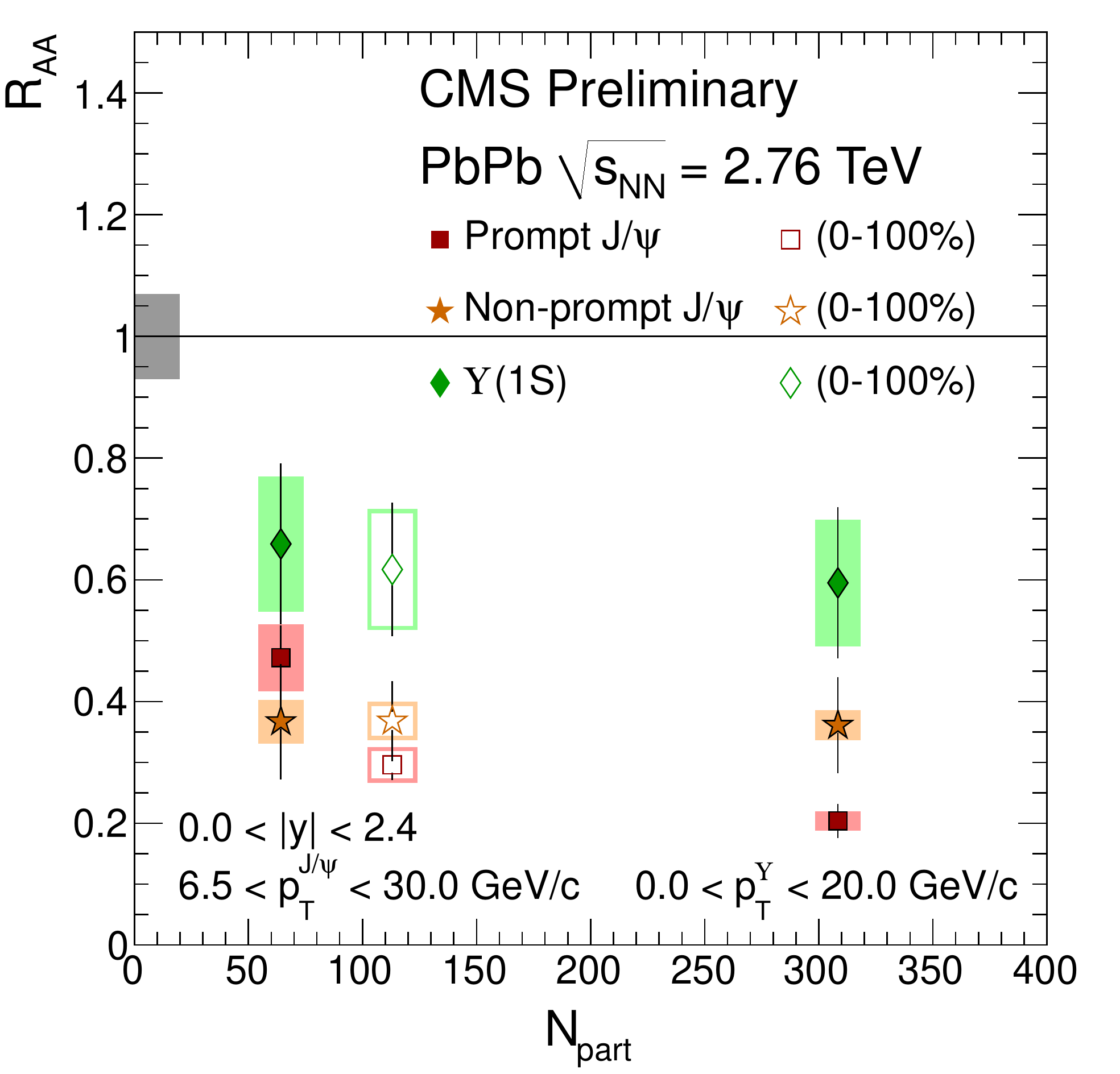}
    \includegraphics[width=0.55\linewidth]{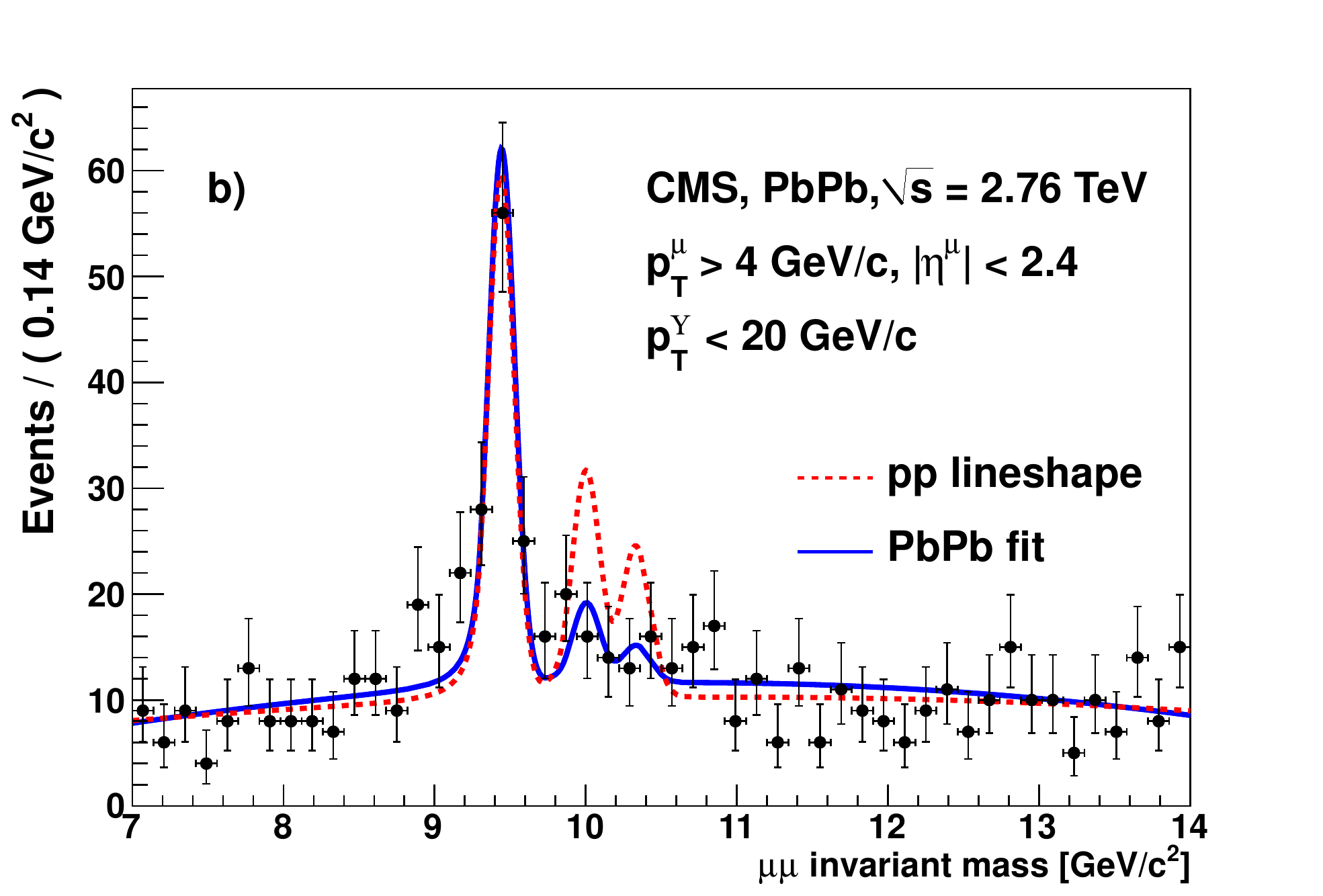}
    \caption{Left: Nuclear modification factors of prompt $J/\psi$ (red squares), non-prompt $J/\psi$ (orange stars) and $\Upsilon(1S)$ (green diamonds) versus $N_{part}$ in two centrality bins ($0-20$ and $20-100$\%, closed symbols) and integrated over centrality (open symbols). Right: dimuon invariant mass plot in PbPb collisions, superimposed to a fit to the corresponding pp data (dashed red line).}
    \label{fig:upsilon}
  \end{center}
\end{figure}

Higher energy and luminosity provided by the LHC allow a copious production of $b$ quarks and thus widen the possibility to study the bottomonium family. CMS reported a preliminary study~\cite{CMSQuarkonia} of the $\Upsilon(1S)$ nuclear modification factor of $R_{AA}=0.62 \pm 0.11 \pm 0.10$ averaged over all centralities, as reported in the left panel of Fig.~\ref{fig:upsilon}, with diamond symbols. The dependencies on event centrality, $\Upsilon$ $p_T$ and rapidity are investigated, within limited statistics. No centrality dependence is observed and at least the mid-rapidity and low $p_T$ $\Upsilon(1S)$ are suppressed, see details in Ref.~\cite{CMSQuarkonia}

The $\Upsilon$ excited states are also searched for. Though visible in the pp data (the red dashed line on the right panel of Fig.~\ref{fig:upsilon} is a fit to pp data) they do not appear above the background in PbPb data (the points, also fitted to the blue curve), in this kinematical range (single muons have $p_T>4$~GeV/$c$). From this, the double ratio of excited to ground states, in PbPb and pp is computed to be:
\begin{equation}\label{eq:upsilon}
\frac{\Upsilon(2S+3S)/\Upsilon(1S)|_{PbPb}}{\Upsilon(2S+3S)/\Upsilon(1S)|_{pp}} = 0.31^{+0.19}_{-0.15}\pm 0.03
\end{equation}
The probability to obtain this measured double ratio (or below) if the $\Upsilon$ excited states are not suppressed is less than 1\%, providing a strong indication for their suppression. This result is published in Ref.\cite{CMSUpsilon}. Note that a large part of the $\Upsilon(1S)$ yield is coming from higher bottomonium states, for instance $\simeq$~50\% mostly from $\chi_b$ for $p_T>8$~GeV/$c$ as measured by CDF in Ref.~\cite{CDF}.

\section{Non-prompt $J/\psi$}

As mentioned above, the CMS collaboration was able to separate prompt from non-prompt $J/\psi$. The latter are quite suppressed as can be seen from the star symbols on the left part of Fig.~\ref{fig:upsilon}. This suppression does not arise from quarkonium interactions, since $J/\psi$ are product of decays that occur long after the medium vanished. Instead, they must reflect the energy loss of $b$-quarks through the medium.

\section{Conclusions}

Exciting new results on quarkonia are coming up from the LHC. We saw that $\Upsilon$ excited states are very likely to be suppressed and that the $\Upsilon(1S)$ ground state is suppressed by about 40\% which could just reflect the higher state feeddown disappearance. At LHC, low and high $p_T$  $J/\psi$ are respectively less and more suppressed than at RHIC. This upside-down behaviour could be the consequence of uncorrelated $c$ and $\overline{c}$ quark recombination. Non-prompt $J/\psi$ are strongly suppressed, which should be a consequence of $b$-quark energy loss. In particular in the $\Upsilon$ case, the uncertainties are large and more PbPb data is needed. In all cases, the behaviour of the above-mentioned observables in a pA-like run will help in establishing a firm interpretation.

\section*{Acknowledgements}
The author presenting these results received funding from the European Research Council under the FP7 Grant Agreement no. 259612.

%\appendix
%\section{First Appendix} %Empty argument \section{} yields `Appendix'.
%\section{Second Appendix}

\end{document}